\documentclass{appolb}
\usepackage{graphicx}
\usepackage{amsmath}
\usepackage{xcolor}
 \usepackage{amssymb} 

\begin{document}
\title{A note on analytic continuation to Minkowski space relevant for $^3P_0$ decay--model phenomenology.
\thanks{Presented at the Excited QCD 2026 Workshop, Granada, Spain.}%
}
\author{
{Alexandre Salas-Bern\'ardez\thanks{Speaker} and Juan Ferrera
\address{Dept. An\'alisis Matem\'atico y Matem\'atica Aplicada \& IPARCOS,
} }\\[3mm]
{Felipe J. Llanes-Estrada  
\address{Dept. F\'isica Te\'orica \& IPARCOS,\\
U. Complutense de Madrid, Plaza de las Ciencias 1 and 3, 28040  Spain}
}\\[3mm]
}
\maketitle
\begin{abstract}
We discuss some aspects of the uniqueness of analytically continuing Euclidean Green functions obtained by means of the Dyson– Schwinger equations (DSEs) and constrained by lattice data, back to Minkowski-space with physical $p^2$.



\end{abstract}

\section{Extending Euclidean Solutions to the Minkowski regime}
Traditionally, computations of fundamental Green's functions of Quantum Chromodynamics (QCD) from its Dyson-Schwinger equations (DSE)~\cite{Roberts:1994dr,Alkofer:2000wg,Alkofer:2004it,Fischer:2006ub} have been performed in the Euclidean regime.  This trend has continued in more recent work~\cite{Pelaez:2017bhh,Huber:2020keu,Aguilar:2024fen,Aguilar:2024ciu} and is common to other approaches such as Lattice gauge theory~\cite{Oliveira:2016muq,Oliveira:2018ukh,Colaco:2024gmt}. Approximations to these functions directly from Minkowski kinematics do exist, such as via Covariant Spectator Theory ~\cite{Biernat:2018khd} or dispersive representations~\cite{Sauli:2020dmx}.

In recent work~\cite{Alkofer:2023syz}, fits to the Euclidean DSE/lattice quark-gluon vertex $\Gamma^\mu$ have been employed to explain how decay matrix elements result in $^3P_0$ quark-antiquark quantum numbers, at low momentum (at higher one the chiral $^3S_1$ structure is found instead). The vertex therein is computed for positive Euclidean variables $p_E^2$ (generically, a complex-plane variable $z$) 
and the decay matrix element is necessary for physical, on-shell momentum $p^2=m_q^2$, which corresponds to specific points along the real, negative $p_E^2=-p^2$ half axis upon inverting the Wick rotation.
We here explore the uniqueness of such analytic continuation and some of its uncertainties.

\section{Ambiguity in the analytic continuation}
We aim to expose the important ambiguity inherent in analytically continuing a function that is known only at a finite set of points. But first, we remind the reader that the analytic continuation is possible.

\subsection{Extending a finite data set}
The discrete data points are typically lattice calculations or numerical solutions to truncated Dyson-Schwinger Equations (DSE), but in other circumstances one can also think of extending   experimental measurements.
To keep the discussion generic, we will denote the sampled values of the independent variable $p_E^2$ by $a_n$, and the Green function values. the images, by $A_n=f(a_n)$ for $n=1,...,k$. A direct way of extending these $(a_n\neq 0,A_n)$ pairs to an analytic function is to construct the Weierstrass polynomial and its derivative,
\begin{equation}
W(z)=\prod_{n=1}^k\left(1-\frac{z}{a_n}\right)\ , \  \ \ \
W'(z)=\sum_{m=1}^k -\frac{1}{a_m}\prod_{n\neq m}\left(1-\frac{z}{a_n}\right)\ :
\end{equation}
$W(z)$ clearly has zeroes at $z=a_n$  $\forall n$. 
Upon substituting $z=a_p$ for a given $p$ in the derivative, all terms in the sum are zero but one, that with $m=p$,
\begin{equation}
W'(a_p)=-\frac{1}{a_p}\prod_{n\neq p}\left(1-\frac{ a_p
} {a_n}\right).
\end{equation}

The Weierstrass polynomial allows to construct an entire function, $F(z)$, whose image contains the wanted values, $F(a_n)=f(a_n)=A_n$  $\forall n$. For this, a denominator $z-a_n$ is used to compensate the zero of the Weierstrass polynomial at the given point $a_n$; $1/W'$ is used to cancel the remaining monomials, and the factors $A_n$ then set the correct function values,
\begin{align}
F_{\{A_n\}}(z)=\sum_{n=1}^k \frac{W(z)}{z-a_n} \frac{A_n}{W'(a_n)}
\end{align}

This function extends the data to the entire complex plane analytically.

\subsection{The extension is not unique}
To expose the ambiguity of any analytic extension, it is enough to write down 
a more general function family, parametrized by $\gamma_n\in \mathbb{R}$, satisfying the same conditions (setting $z=p_E^2$, the Euclidean momentum),
\begin{align}
F_{\{\gamma_n\}}(p_E^2)= 
\sum_{n=1}^k e^{\gamma_n(p_E^2-a_n)} \frac{W(p_E^2)}{p_E^2-a_n} \frac{A_n}{W'(a_n)},
\label{NonUniqueF} \ .
\end{align} 
(The addition of a small positive imaginary part to all the parameters, $\gamma_n\to \gamma_n+i\epsilon$ forces $F_{\gamma_n}\to 0$ for $|p_E^2|\to \infty$ in the upper half plane, but it has an essential singularity at infinity nonetheless. It does not yield a tempered distribution,  so it would not appear in a local field theory.) 

\begin{figure}[ht!]
    \centering
\includegraphics[width=0.49\columnwidth]{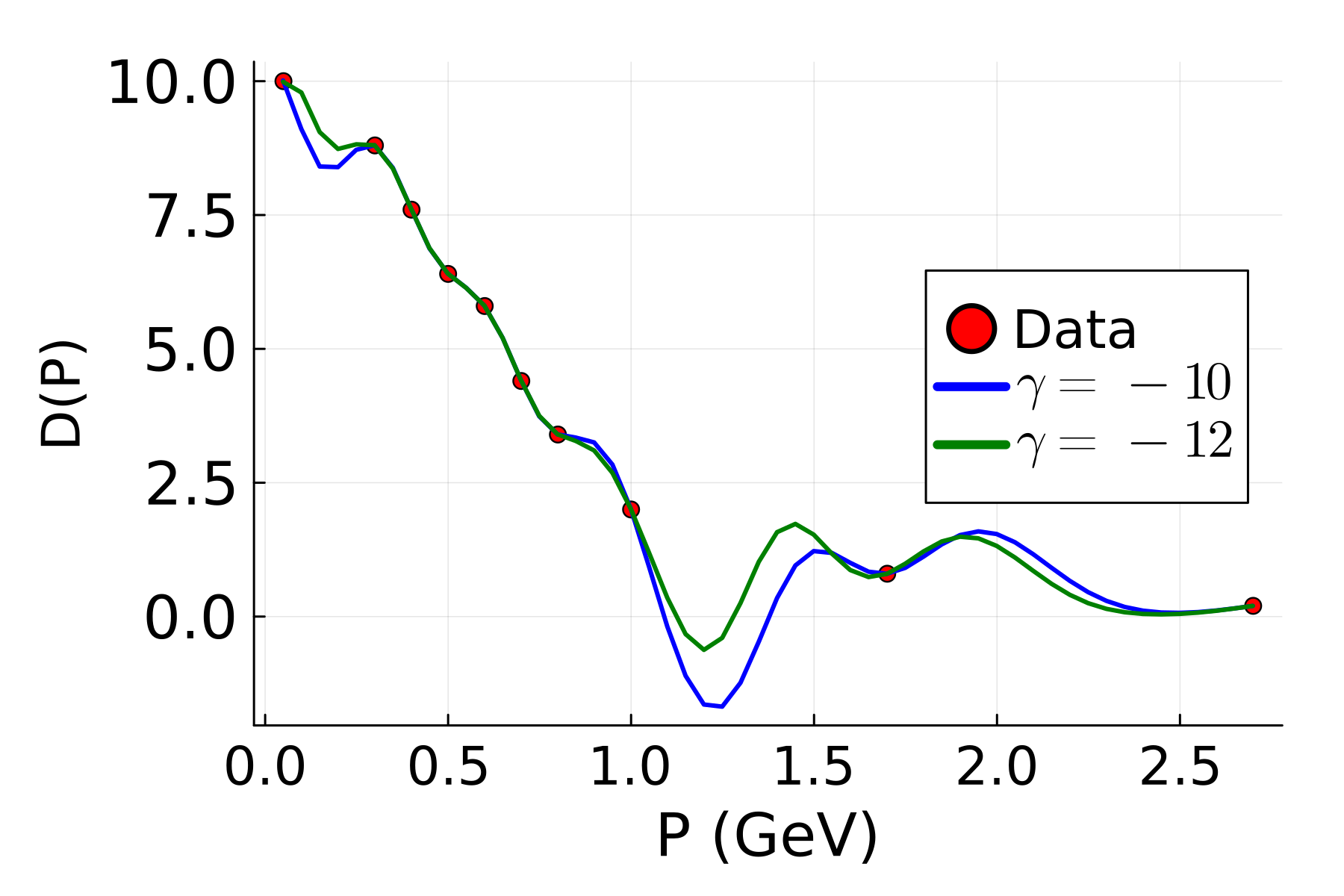}
\includegraphics[width=0.40\linewidth]{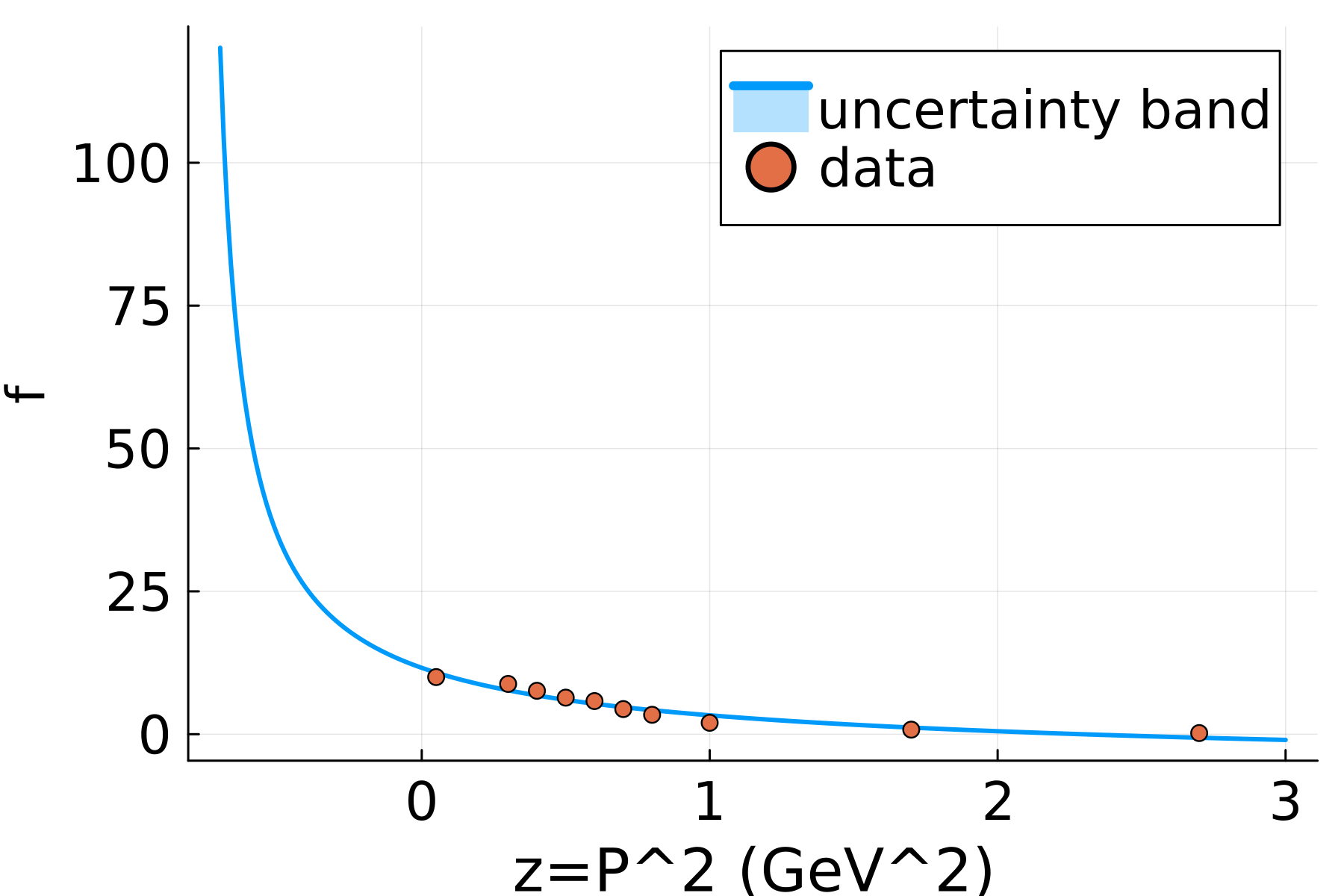}
    \caption{{\bf Left: }Circles (red online): small sample of the gluon propagator data, taken as  a function $f(x)$ whose analytic extension is wanted. Lines: two possible analytic functions passing by all points from the Weierstrass construction
    in Eq.~(\ref{NonUniqueF})
    {\bf Right:} Example analytic continuation of the same data of fig.~\ref{fig:ambiguous} with one pole and one log cut on the left axis, imposing monotonicity.
    }
    \label{fig:ambiguous}
\end{figure}

Figure \ref{fig:ambiguous} shows an example set of points and two plausible functions $F_{\gamma_n}$ which via this Eq.~(\ref{NonUniqueF}) (except with a Gaussian $e^{\gamma(p_E^2-a_n)^2}$ instead of a simple exponential) can be used to extend them, with respective parameters $\gamma_n=-10$ (blue online) and $\gamma_n=-12$ (green online). 

\color{black}

The figure clearly illustrates that, since numerically represented functions are not known over a complete segment, the analytic extension is not unambiguous without additional hypothesis. For example, to discard functions with unwanted oscillation, we can impose the condition of monotonicity between any two consecutive pairs of points.

\subsection{Adding singularities}
Titchmarsh's theorem applied to causal Green's functions~\cite{Llanes-Estrada:2019ktp} forbids poles on the energy upper-half  plane $p^0=E+i\epsilon$,  $\epsilon>0$. Upon folding to the Lorentz-invariant  
$p^2=(p^0)^2-{\bf p}^2=(E^2-\epsilon^2) - {\bf p}^2 +2iE\epsilon$,
because the (real part of the) energy $E$ may have either sign,
the first Riemann sheet of $p^2$ must be singularity free, except on its real, positive axis (which becomes the negative one upon swapping to Euclidean momentum $p^0_E=ip^0$). This means that, a priori, there is no obstruction to analytically continuing over the first Riemann sheet to the left $p_E^2<0$ half-axis, but singularities could appear there (and also in other Riemann sheets if the function is cut).

The sought function is now assumed to be analytic in the extended domain, but to maybe present poles outside that region of interest, say for very negative real $p_E^2<0$. These poles can be incorporated as follows, to yield a nonunique
meromorphic function $f(z)$. The poles may be isolated (in one dimension), or eventually,  form surfaces in $\mathbb{C}^3$ (each generated by, say for the quark--gluon vertex $\Gamma(p_{1E}^2,p_{2E}^2,p_{gE}^2)$, finding a pole in one variable $p_{1E}^2$ and then varying its position upon displacing $p_{2E}^2$ and $p_{gE}^2$).

To incorporate poles of $F$
at given locations (for example, the CDD poles~\cite{Castillejo:1955ed} of scattering theory), the construction of Eq.~(\ref{NonUniqueF}) can be modified to fix the values of the function to the ``data'' points while adding a pole of order $N$ at $z=z^\ast$ simply by
\begin{align}
F_{\{\gamma_n\},z^*}(p_E^2)= 
\sum_{n=1}^k e^{\gamma_n(p_E^2-a_n)} \frac{W(p_E^2)}{p_E^2-a_n} \frac{A_n}{W'(a_n)}+\frac{W(p_E^2)}{(p_E^2-(p_E^2)^\ast)^N}\ ;
\label{NonUniqueF2}
\end{align}
The function at the given data points is unaffected because   $W(a_n)=0$.
The extension to a function which displays a cut is now obvious, for example,
\begin{align}
\tilde{F}_{\{\gamma_n\}}(p_E^2)= 
\sum_{n=1}^k e^{\gamma_n(p_E^2-a_n)} \frac{W(p_E^2)}{p_E^2-a_n} \frac{A_n}{W'(a_n)}+{W(p_E^2)}\log(p_E^2-(p_E^2)^*)\ .
\label{NonUniqueF3}
\end{align}
\section{The asymptotic behaviour is insufficient to lift the ambiguity}

In this section, we dispel the idea that the asymptotic behaviour of a Green function at large Euclidean momentum—known from perturbation theory—can uniquely determine its analytic continuation.

For example, suppose that a behaviour of the form $(p_E^2)^{-K}$ at infinity is desired. One might attempt to modify Eq.~(6) by introducing a multiplicative factor depending on a parameter $a \in \mathbb{R}$,
\begin{equation}
F_{\{\gamma_n\}}(p_E^2) =
\left[
\sum_{n=1}^{k} 
e^{\gamma_n (p_E^2 - a_n)} 
\frac{W(p_E^2)}{p_E^2 - a_n} 
\frac{A'_n}{W'(a_n)}
\right]
(p_E^4 + a^2)^{-k/2 - K/2}.
\end{equation}

(This choice preserves the reality of the function along the real axis, while introducing additional poles and branch cuts at $p_E^2 = \pm i a$; these, however, are not relevant for the present argument.)

The values at the data points must then be adjusted through suitable coefficients $A'_n$. As an illustration, consider a mass function satisfying the anomalous-dimension equation along the negative real axis,
\begin{equation}
K = - \frac{p_E^2}{M(p_E^2)} \frac{dM}{dp_E^2},
\end{equation}
with $K = \frac{3C_F}{(4\pi)^2}$, where $C_F$ is the fundamental Casimir.

We now invoke Arakelyan’s theorem to show that, even when both the data and the asymptotic behaviour are fixed, the analytic continuation remains non-unique. The theorem states:

Let $\Omega \subset \mathbb{C}$ be an open set and $E \subset \Omega$ a relatively closed subset, with interior $E^\circ$. Then, for every function $f$ that is continuous on $E$ and holomorphic on $E^\circ$, and for every $\varepsilon > 0$, there exists a holomorphic function $\tilde{g} : \Omega \to \mathbb{C}$ such that
\begin{equation}
|f(z) - \tilde{g}(z)| < \varepsilon \quad \text{for all } z \in E,
\end{equation}
if and only if the Alexandroff compactification $\Omega^*$ is connected and locally connected.

The case $\Omega = \mathbb{C}$ and $E = \mathbb{R}$ is particularly relevant, since $\mathbb{C}^* \setminus \mathbb{R}$ is connected and locally connected.

Suppose we seek an entire function $g$ that takes real values on $\mathbb{R}$, matches a finite set of (numerically determined) data points $\{a_1, \dots, a_N\}$, and satisfies a prescribed asymptotic condition such as
\begin{equation}
\lim_{x \to +\infty} \frac{g(x)}{x^2} = 1.
\end{equation}

To demonstrate the ambiguity, construct an auxiliary continuous function $f : \mathbb{R} \to \mathbb{R}$ such that: $f(a_k) = g(a_k)$ for all data points, $f(x)$ shares the same asymptotic behaviour, but $f$ differs from $g$ at some point $x_0 \in \mathbb{R}$, with $|f(x_0) - g(x_0)| = 1 + \varepsilon$.
Such a function is straightforward to construct.

By Arakelyan’s theorem, there exists an entire function $\tilde{g}$ approximating $f$ on $\mathbb{R}$,
\begin{equation}
|f(z) - \tilde{g}(z)| < \varepsilon,
\end{equation}
while still satisfying the asymptotic condition and reproducing the data within numerical accuracy, $\tilde{g}(a_k) \approx g(a_k)$.

Choosing $\varepsilon$ smaller than the numerical uncertainty in the data allows one to enforce exact interpolation, $\tilde{g}(a_k) = g(a_k)$, without affecting the argument. We thus obtain two distinct functions $g$ and $\tilde{g}$ that share the same asymptotic behaviour, reproduce the same data points, yet differ by a finite amount at some point $x_0$.

\section{Commentary}

As a consequence of the discussion on the ambiguity, the various methods employed in searching for adequate analytic extensions, such as
Pad\'e Approximants (including Schlessinger's Point Method),
direct numerical solutions of the differential Cauchy Riemann Equations, or the extension via a K\"allen-Lehmann representation (a Stieltjes transform) entail ambiguity which manifests itself in various forms depending on the chosen method.

We thus need to swipe the space of functions passing by the data and satisfying certain reasonable conditions, and extrapolating all the functions of this \emph{fascis}, generate different possible analytic extrapolations to the Minkowski point which provide us with an uncertainty band on the extrapolated value. For example, imposing the additional requirements of monotonicity, one particle pole and one cut (right plot of figure~\ref{fig:ambiguous})
one can obtain a ``cleaner'' extrapolation. 
\section*{Acknowledgments}
Supported by grants 
PID 2022-137003NB-I00  and  2023-148162NB-C2
of the Spanish MCIN/AEI/10.13039/501100011033/ and MICIU.

\end{document}